\documentclass{article}%
\usepackage{amsmath}
\usepackage{amsfonts}%
\usepackage{amssymb}

\begin{document}

\title{Novel approach to the study of 
\\quantum effects in the early universe}
\author{A. Geralico$^{1,2,3}$\thanks{ e-mail: Andrea.Geralico@le.infn.it}, G.
Landolfi$^{1,2}$\thanks{ e-mail: Giulio.Landolfi@le.infn.it}, G.
Ruggeri$^{1,2}$\thanks{ e-mail: Giovanna.Ruggeri@le.infn.it} and G.
Soliani$^{1,2}$\thanks{ e-mail: Giulio.Soliani@le.infn.it}
\\$^{1}$\emph{Dipartimento di Fisica dell'Universit\`{a} di Lecce, I-73100
Lecce, Italy}\\$^{2}$\emph{INFN, Sezione di Lecce, I-73100 Lecce, Italy} 
\\$^{3}$\emph{International Center for Relativistic Astrophysics-I.C.R.A., }
\\\emph{Universit\`{a} di Roma ''La Sapienza'', I-00185 Roma, Italy}}
\date{}
\maketitle

\begin{abstract}

We develop a theoretical frame for the study of classical and 
quantum gravitational waves based on the properties of a 
nonlinear ordinary differential equation for a 
function $\sigma(\eta)$ of the conformal time $\eta$, 
called the auxiliary field equation. 
At the classical level, $\sigma(\eta)$ 
can be expressed by means of two independent 
solutions of the ''master equation'' to which 
the perturbed Einstein equations for 
the gravitational waves can be reduced. 

At the quantum level, all the significant physical quantities 
can be formulated using Bogolubov transformations and the operator quadratic 
Hamiltonian corresponding to the classical version of a damped parametrically 
excited oscillator where the varying mass is replaced by the square 
cosmological scale factor $a^{2}(\eta)$. 
A quantum approach to the generation of gravitational 
waves is proposed on the grounds of the previous $\eta-$dependent Hamiltonian.
An estimate in terms of $\sigma(\eta)$ and $a(\eta)$ of the destruction of 
quantum coherence due to the gravitational evolution and an exact expression 
for the phase of a gravitational wave corresponding to any value of $\eta$ 
are also obtained. 
We conclude by discussing a few applications to quasi-de Sitter and standard 
de Sitter scenarios.

\hfill




\end{abstract}

\clearpage

\section{Introduction}

\setcounter{equation}{0}

In a notable paper by Grishchuk and Sidorov [\ref{grishchuk}], it was shown
that relic gravitons can be created from the vacuum quantum fluctuations of
the gravitational field during cosmological expansion and can be
interpreted as squeezed quantum states of the 
gravitational field, in analogy to what happens
for squeezed states in quantum optics [\ref{stoler}-\ref{schumaker}]. A
systematic treatment of the particle creation mechanism is contained in
[\ref{birrel}]. The theory of particle creation is essentially based on the
Bogolubov transformations, exploited primary in the formulation of the process
of squeezing for the electromagnetic field.
An important aspect of the theory of particle creation is the possible loss of
coherence in quantum gravitational theories. In a sense, the exploration of
decoherence in gravitational theories has been more concerned with the quantum
(gravity) effects which can manifest themselves even at scales others than the
Planck one. The growing interest in this field relies, among the others, on
quite a remarkable mechanism: the amplification of quantum (incidentally
vacuum) fluctuations in the metric of a gravitational background (see
\textit{e.g.} [\ref{grishchuk},\ref{grishchuk2}-\ref{maggiore}]). In the
presence of a change of regime under the cosmological evolution, the
occupation number of the initial quantum state would get indeed amplified. As
long as the change can be considered as adiabatic, the amplification factor
approaches one. Nevertheless, in case the change is sudden the
amplification mechanism cannot be neglected. In such a case, even the vacuum
state transforms into a multiparticle state in the Fock space appropriate to
the new regime. The scenario clearly sounds highly attractive under the
stronger convintion that, thanks to the great progress we are witnessing in
the experimental application of new technologies, the amplification mechanism
may provide the possibility to detect quantum effects (\textit{e.g.} relic
gravitons) at scales considerably above the Planck one.
Because of the semiclassical approximation underlying these studies, 
the natural
formal arena turns out to be that of coherent states. These states are
generated by the displacement operator $D\left(  \alpha\right)  $ 
(see Section 2). 
Squeezed states enter in the matter whenever the quadratic operators
$\hat{a}^{2}$ and $\hat{a}^{\dagger\,\,2}$ are involved. A squeezed state is generated by
the action on a coherent state of the so-called squeeze operator defined in
Section 2.
As a matter of fact, in a Friedmann-Robertson-Walker (FRW) spacetime the
behavior of matter scalar fields as well as of gravitational waves is governed
by an equation of the time-dependent type (or time-dependent oscillator, TDO).
Thus the problem of both particle creation and metric field fluctuation
amplification during the cosmological evolution is reduced to that of solving
the quantum TDO problem. The latter problem has been the subject of several
studies, mainly in connection with quantum optical arguments.
In this paper, the main idea is to make use of the machinery built-up in
[\ref{profilo}] and further developed in [\ref{landolfi}]. The formalism would
enable us to find at every stage information concerning the spectrum of
created modes. In the case of a de Sitter gravitational field, the
prescription straightforwardly results into full determination of quantities
of physical relevance, such as Bogolubov coefficients
and the phase of gravitational waves for any value of the conformal time
$\eta$. All these quantities are determined exactly. This has been possible by
solving a nonlinear ordinary differential equation for an auxiliary field
$\sigma(\eta)$ which has been expressed in terms of two independent solutions
of the $\eta-$dependent part of the D'Alembert equation for the gravitational
perturbation tensor field.
Another interesting result achieved in this paper is an estimate in terms of
the auxiliary field $\sigma(\eta)$ and the scale factor $a(\eta)$ of the
destruction of quantum coherence due to gravitational time-evolution.
In our approach, the use of the (nonlinear) auxiliary equation in the linear
parametrically excited oscillator equation for $y(\eta)$ (see Eq.
(\ref{tdoy})) reveals therefore to be quite profitable and makes more
strict and compact the formal connection between quantum optics and the
theory of gravitational waves.
The paper is organized as follows. In Section 2, after a description of
some basic properties pertinent to the squeeze operator (\ref{squeeze}),
we introduce the nonlinear equation (\ref{eqausiliaria}) for the
auxiliary field $\sigma(t)$ in terms of which the position and momentum
operators $Q$ and $P$ turn out to be expressed. The role of the matrix
element between squeezed states $\left|  \alpha,z\right\rangle $ of the
operator $D(\alpha)\,S(z)\,H(t)\,S^{\dagger}(z)\,D^{\dagger}(\alpha)$ is
investigated. This matrix element results to be evaluated in terms of the
auxiliary field $\sigma(t)$. It is worth noticing that in the expression for
the matrix element three energy terms appear, one of them, formula
(\ref{10.2}), can be interpreted as the energy related to squeezed states
which do not preserve the minimum uncertainty. In the theory of gravitational
waves, Eq. (\ref{10.2}) plays the role of \textit{decoherence} energy of the
waves. The Bogolubov transformation is reported whose coefficients are 
explicitly written in terms of $\sigma$.
In Section 3 we discuss classical and quantum aspects of the generation of
gravitational waves. An exact formula for the phase of a gravitational wave
is obtained. A natural approach to the theory of gravitational waves based 
on the Kanai-Caldirola oscillator is outlined by means of an operator 
Hamiltonian expressed in terms of the auxiliary field $\sigma(\eta)$.
 In Section 4 some applications are displayed. Precisely, we evaluate the 
decoherence energy in the quasi-de Sitter inflationary model and standard de
Sitter spacetime and the role of Bogolubov coefficients in terms of the
auxiliary field in the particle creation mechanism is analyzed. Finally, in
Section 5 some future perspectives are discussed.

\section{Preliminaries on the squeezed states in generalized oscillators}
\setcounter{equation}{0}

Let us recall that a squeezed state of a quantum system is defined by
[\ref{stoler}]%
\begin{equation}
\left|  \alpha,z\right\rangle =D(\alpha)\,S(z)\,\left|  0\right\rangle
\label{2.1}%
\end{equation}
where%
\begin{equation}
D\left(  \alpha\right)  =e^{\alpha\hat{a}_{0}^{\dagger}-\alpha^{\ast
}\hat{a}_{0}}\label{weyl}%
\end{equation}
is the (unitary) displacement (Weyl) operator,
\begin{equation}
S(z)=e^{\frac{1}{2}z\hat{a}_{0}^{\dagger\,\,2}-\frac{1}{2}z^{\ast}%
\hat{a}_{0}^{2}}\label{squeeze}%
\end{equation}
is the (unitary) \textit{squeeze} operator [\ref{stoler},\ref{hollerhorst}],
$\hat{a}_{0}$ and $\hat{a}_{0}^{\dagger}$ stand for $\hat{a}(t)|_{t=t_{0}}$
and $\hat{a}^{\dagger}(t)|_{t=t_{0}}$, respectively, 
where $\hat{a}(t)$ and $\hat{a}^{\dagger}(t)$ 
denote the annihilation and creation operators of the system. The
complex functions $\alpha(t)$ and $z(t)$ are arbitrary, namely
\begin{equation}
\alpha=\left|  \alpha\right|  \,e^{i\varphi}\;,\qquad z=r\,e^{i\phi
}\;,\label{defr}%
\end{equation}
with $\varphi$, $\phi$ arbitrary (real) $c$-numbers. Notice that for $z=0$,
Eq. (\ref{2.1}) reproduces the coherent state $\left|  \alpha,0\right\rangle
=D(\alpha)\,\,\left|  0\right\rangle $.

The following relations
\begin{align}
b  &  \doteqdot S^{\dagger}\hat{a}_{0}S=\hat{a}_{0}\cosh r+\hat{a}_{0}%
^{\dagger}e^{i\phi}\sinh r\label{3.1}\\
b^{\dagger}  &  \doteqdot S^{\dagger}\hat{a}_{0}^{\dagger}S=\hat{a}_{0}%
^{\dagger}\cosh r+\hat{a}_{0}e^{-i\phi}\sinh r \label{3.2}%
\end{align}
hold. This can be readily seen by applying the Baker-Campbell-Hausdorff
formula.
In other words, the squeeze operator $S(z)$ induces a canonical transformation
of the annihilation and creation operators, 
in the sense that $\left[b,b^{\dagger}\right] =\hat{1}$.

\subsection{Transformation of the position and momentum variables under the
squeeze operator $S(z)$}

From the quantum theory of generalized oscillators, it follows that position
and momentum operators $Q$ and $P$ can be expressed by [\ref{profilo}%
,\ref{landolfi}]%
\begin{equation}
Q=\sqrt{\frac{\hbar}{m}}\,\sigma\text{ }\left(  \hat{a}+\hat{a}^{\dagger
}\right)  \;,\qquad P=\sqrt{\hbar m}\left(  \xi\hat{a}+\xi^{\ast
}\hat{a}^{\dagger}\right)  \;, \label{4.2}%
\end{equation}
with $\hat{a}=\hat{a}(t)$, $\hat{a}^{\dagger}=\hat{a}^{\dagger}(t)$ and
\begin{equation}
\xi=\frac{-i}{2\sigma}+\left(  \dot{\sigma}-\frac{M}{2}\sigma\right)  \;,
\label{defxi}%
\end{equation}
where $M=M(t)\equiv\frac{\dot{m}}{m}$, the dot means time derivative, and (the
mass) $m=m(t)$ is a given functions of time. The function (c-number)
$\sigma(t)$ satisfies the nonlinear ordinary differential equation
[\ref{ermakov},\ref{pinney}]%
\begin{equation}
\ddot{\sigma}+\Omega^{2}\sigma=\frac{1}{4\sigma^{3}}\,\,\,,
\label{eqausiliaria}%
\end{equation}
($\Omega$ is specified below, see Eq. (\ref{defOmega2})), called the 
auxiliary equation associated with the classical equation of motion
\begin{equation}
\ddot{q}+M\dot{q}+\omega^{2}(t)\,q=0\;, \label{eqtdo}%
\end{equation}
and $\omega(t)$ is the time dependent frequency. Via the transformation
\begin{equation}
q\rightarrow e^{-\frac{1}{2}\int_{t_{0}}^{t}M(t^{\prime})\,dt^{\prime}%
}\,y\qquad, \label{qqtilde}%
\end{equation}
Eq. (\ref{eqtdo}) can be cast into the equation%
\begin{equation}
\ddot{y}+\Omega^{2}(t)\,y=0\qquad, \label{tdoy}%
\end{equation}
where%
\begin{equation}
\Omega^{2}(t)=\frac{1}{4}\left(  4\omega^{2}-2\dot{M}-M^{2}\right)  \qquad.
\label{defOmega2}%
\end{equation}
The quantum theory of the generalized oscillator (\ref{eqtdo}) can be
described by the Hamiltonian operator [\ref{lewis},\ref{profilo}]%
\begin{equation}
H(t)=\frac{P^{2}}{2m}+\frac{1}{2}m\,\omega^{2}Q^{2} \label{5.5}%
\end{equation}
where the canonical variables $Q,P$ are given by Eq. (\ref{4.2}).

Taking account of Eqs. 
(\ref{3.1}) and (\ref{3.2}), and choosing $\phi=0$ we
obtain\footnote{Stoler [\ref{stoler}] saw that, generally, states of the type
$\left|  \alpha,z\right\rangle $ do not describe wave packets relative to the
minimum value of the product $\left(  \Delta Q\right)  \,\left(  \Delta
P\right)  $, where $\Delta$ means the variance operation (see Eq.
(\ref{6.4})). The state $\left|  \alpha,z\right\rangle $ can describe a wave
packet of minimum uncertainty only if $z$ is real ($\phi=0$). In the framework
of quantum generalized oscillators, this corresponds to the condition
(\ref{6.3}).}%
\begin{align}
S^{\dagger}QS  &  = e^{r}Q\;,\label{6.1}%
\\
S^{\dagger}PS  &  
=2m\sigma\left(  \dot{\sigma}-\frac{M}{2}\sigma\right)  \sinh r
\, Q+e^{-r}P\;. \label{6.2}%
\end{align}
The physical meaning of the expression $\dot{\sigma}-\frac{M}{2}\sigma$ will
be clarified later. At the present we observe only that whenever the
condition
\begin{equation}
\dot{\sigma}-\frac{M}{2}\sigma=0 \label{6.3}%
\end{equation}
is fulfilled, then the uncertainty product $\left(  \Delta Q\right)  \,\left(
\Delta P\right)  \,\ $of the variances%
\begin{equation}
\left(  \Delta Q\right)  =\sqrt{\left\langle Q^{2}\right\rangle -\left\langle
Q\right\rangle ^{2}}\;,\qquad\left(  \Delta P\right)  =\sqrt{\left\langle
P^{2}\right\rangle -\left\langle P\right\rangle ^{2}}\;, \label{6.4}%
\end{equation}
attains its minimum, where the expectation value $\left\langle \dots
\right\rangle $ is referred to coherent states [\ref{landolfi}].

We observe that the operators $S^{\dagger}QS$ and $S^{\dagger}PS$ obey the
same commutation relation as $Q$ and $P$, that is%
\begin{equation}
\left[  S^{\dagger}QS,S^{\dagger}PS\right]  =\left[  Q,P\right]  =i\hbar\;.
\end{equation}

However, in contrast to what happens for the operators $b$ and $b^{\dagger}$
(see Eqs.
(\ref{3.1})-(\ref{3.2})), the operators $S^{\dagger}QS$ and $S^{\dagger
}PS$ are not Hermitian conjugate. We point out that the property of Hermitian
conjugation is enjoyed by the operators $S^{\dagger}QS$ and $S^{\dagger}PS$ in
the case in which the condition (\ref{6.3}) is valid.

A possible physical interpretation of the properties (\ref{6.1}) and
(\ref{6.2}) is the following. For a quantum system governed by a Hamiltonian
preserving the minimum wave packet (\textit{i.e.}, the condition (\ref{6.3})
holds) Eqs. (\ref{6.1}) and (\ref{6.2}) become
  $ S^{\dagger}QS =e^{r}Q $ and $ S^{\dagger}PS  =e^{-r}P$,  respectively. 
If $\left|  \psi\right\rangle $ is the state of the system under
consideration, then $\left|  \psi^{\prime}\right\rangle =S(r)\,\left|
\psi\right\rangle $ represents the same system squeezed in the space of the
position $Q$ by a factor $e^{-r}$ and expanded in the space of the momentum
$P$ by the factor $e^{r}$. In fact, we deduce%
\begin{align}
e^{-r}\left\langle \psi^{\prime}\left|  \,Q\,\right|  \psi^{\prime
}\right\rangle  &  =\left\langle \psi\left|  \,Q\,\right|  \psi\right\rangle
\quad , \quad 
e^{r}\left\langle \psi^{\prime}\left|  \,P\,\right|  \psi^{\prime
}\right\rangle  &  =\left\langle \psi\left|  \,P\,\right|  \psi\right\rangle
\quad.
\end{align}

Now, we shall evaluate a matrix element involving the operator
$H(t)$ (see Eq. 
(\ref{5.5})) in the context of squeezing of a quantum system. In
doing so, let us consider the following expectation value between squeezed
states:%
\begin{multline*}
\left\langle \alpha,z\left|  \,D(\alpha)\,S(z)\,H(t)\,S^{\dagger
}(z)\,D^{\dagger}(\alpha)\,\right|  \alpha,z\right\rangle =\\
=\left\langle 0\left|  \,S^{\dagger}(z)\,D^{\dagger}(\alpha) 
D(\alpha)\,S(z)\,H(t)\,S^{\dagger}(z)\,D^{\dagger}(\alpha)D(\alpha)\,S(z)\,\right|
0\right\rangle =\qquad
\end{multline*}%
\begin{equation}
\qquad\qquad\qquad=\left\langle 0\left|  \,H(t)\,\right|  0\right\rangle
=\frac{1}{2m}\left\langle 0\left|  \,P^{2}\,\right|  0\right\rangle +\frac
{1}{2}m\omega^{2}(t)\,\left\langle 0\left|  \,Q^{2}\,\right|  0\right\rangle
\quad\quad, \label{9.1}%
\end{equation}
where Eqs. (\ref{2.1}) and (\ref{5.5}) have been employed.

The expectation values on the right hand side of Eq. 
(\ref{9.1}) can be evaluated
from Eq. (\ref{4.2}). They read%
\begin{align}
\left\langle 0\left|  \,P^{2}\,\right|  0\right\rangle  &  =\hbar\,m\,\left|
\xi\right|  ^{2}=\hbar\,m\left[  \frac{1}{4\sigma^{2}}+\left(  \frac{M}%
{2}\sigma-\dot{\sigma}\right)  ^{2}\right]  \;,\label{9.2}\\
\left\langle 0\left|  \,Q^{2}\,\right|  0\right\rangle  &  =\frac{\hbar}%
{m}\sigma^{2}\;, \label{9.3}%
\end{align}
where $\xi$ and $\sigma$ are described by Eqs. (\ref{defxi}) and 
(\ref{eqausiliaria}). With the help of Eqs. 
(\ref{9.2}) and (\ref{9.3}), Eq.
(\ref{9.1}) takes the form%
\begin{align}
\left\langle \alpha,z\left|  \,D(\alpha)\,S(z)\,H(t)\,S^{\dagger
}(z)\,D^{\dagger}(\alpha)\,\right|  \alpha,z\right\rangle  &  =\left\langle
0\left|  \,H(t)\,\right|  0\right\rangle =\nonumber\\
&  =\left[  \frac{\hbar}{8\sigma^{2}}+\frac{\hbar}{2}\left(  \frac{M}{2}%
\sigma-\dot{\sigma}\right)  ^{2}\right]  \;+\frac{\hbar}{2}\omega
^{2}(t)\,\sigma^{2}. \label{10.1}%
\end{align}
The term in the square bracket corresponds to the vacuum expectation value of
the kinetic energy of the system, while the last term is related to the vacuum
expectation value of the potential energy.

The quantity
\begin{equation}
E_{NM}\doteqdot\frac{\hbar}{2}\left(  \frac{M}{2}\sigma-\dot{\sigma}\right)
^{2}\label{10.2}%
\end{equation}
(NM=non-minimum) can be interpreted as the energy associated with the squeezed
states which do not satisfy the criterium of minimum uncertainty ($E_{NM}%
\neq0$). When the criterium is verified, then $E_{NM}=0$. In such a case
Eq. (\ref{10.1}) can be written as
\begin{align}
\left\langle \alpha,z\left|  \,D(\alpha)\,S(z)\,H(t)\,S^{\dagger
}(z)\,D^{\dagger}(\alpha)\,\right|  \alpha,z\right\rangle  &  =\left\langle
0\left|  \,H(t)\,\right|  0\right\rangle =\nonumber\\
&  =\frac{\hbar}{8\sigma^{2}}\;+\frac{\hbar}{2}\omega^{2}(t)\,\sigma
^{2}.\label{10.3}%
\end{align}
Hence in the minimum uncertainty situation the vacuum expectation values of
the kinetic energy and the potential energy turn out to be proportional to
$\sigma^{2}$ and $\frac{1}{\sigma^{2}}$, respectively. It is noteworthy that,
in general, all the energies appearing in Eq. (\ref{9.1}) can be expressed 
in terms of the auxiliary field $\sigma(t)$ obeying the auxiliary equation
(\ref{eqausiliaria}). This enhances the convenience of our approach to the
study of cosmological quantum effects based on the theory of Eq.
(\ref{eqausiliaria}), which we are going to develop in Section 3.

We remark that the quantity $\left(  \frac{M}{2}\sigma-\dot{\sigma}\right)  $
is connected with the expectation value of the operator $\left\{  Q,P\right\}
=QP+PQ$ between vacuum states, \textit{i.e.}
\begin{equation}
\left\langle 0\left|  QP+PQ\right|  0\right\rangle =2\hbar\sigma\left(
\frac{M}{2}\sigma-\dot{\sigma}\right)  \;. \label{11.1}%
\end{equation}
Thus the minimum uncertainty requirement $\frac{M}{2}\sigma=\dot{\sigma}$
implies that the expectation value $\left\langle 0\left|  QP+PQ\right|
0\right\rangle $ is vanishing.

\subsection{A link between Eq. (\ref{tdoy}) and the auxiliary equation}

For later convenience (see Section 3), we shall report a result establishing a
relationship involving the solutions of the (linear) equation of motion and
the (nonlinear) auxiliary equation
\begin{equation}
\ddot{\sigma}+\Omega^{2}(t)\,\sigma=\frac{\kappa}{\sigma^{3}}\qquad,
\label{eqausiliariak}%
\end{equation}
where $\kappa$ is a constant. If $y_{1}$ and $y_{2}$ are two independent
solutions of Eq. (\ref{tdoy}), then the general solution of the auxiliary
equation (\ref{eqausiliariak}) can be written as [\ref{eliezer}]%

\begin{equation}
\sigma=(Ay_{1}^{2}+By_{2}^{2}+2Cy_{1}y_{2})^{\frac{1}{2}}\qquad,
\label{sigmasqrt}%
\end{equation}
with $A,B,C$ arbitrary constants such that%

\begin{equation}
AB-C^{2}=\frac{\kappa}{W_{0}^{2}} \label{eqABCW0}%
\end{equation}
where $W_{0}=W_{0}(y_{1},y_{2})=y_{1}\dot{y}_{2}-\dot{y}_{1}y_{2}=const$ is
the Wronskian.

It is worth remarking that from the theory of the auxiliary equation
(\ref{eqausiliariak}) a phase can be given by the real function
$\theta(t)$, 
\begin{equation}
\theta(t)=\int_{t_{0}}^{t}\frac{dt^{\prime}}{\sigma^{2}(t^{\prime})}%
\qquad.\label{phasetheta}%
\end{equation}
(See [\ref{ermakov},\ref{pinney},\ref{goff}]; for some applications:
[\ref{lewis},\ref{profilo}].) 
Here we shall suggest the procedure which can be
used to compute the above integral in general cases. To this aim it is
convenient to introduce the function%
\begin{equation}
\psi(t)=\sqrt{A}\,e^{i\alpha}\,y_{1}(t)-\sqrt{B}\,e^{i\beta}\,\,y_{2}(t)\;,
\end{equation}
where $\alpha,\beta$ are real numbers and $y_{1},y_{2}$ are the two
independent solutions appearing in Eq. (\ref{sigmasqrt}). We have
\begin{equation}
\left|  \psi(t)\right|  ^{2}=A\,y_{1}^{2}+B\,y_{2}^{2}-2\sqrt{AB}\,\cos
\theta_{0}\,\,y_{1}\,y\,_{2}\;,\label{psit2}%
\end{equation}
where $\theta_{0}=\alpha-\beta$. Comparing Eq. 
(\ref{psit2}) with Eq. (\ref{sigmasqrt}%
) we get $C=-\sqrt{AB}\,\cos\theta_{0}$, so that the condition
(\ref{eqABCW0}) becomes%
\begin{equation}
AB\sin^{2}\theta_{0}=\frac{\kappa}{W_{0}^2}\;,\label{ABsinkW0}%
\end{equation}
from which $\kappa\neq0$ whenever $\sin\theta_{0}\neq0$. Hence, the auxiliary
field $\sigma$ can be expressed by
\begin{equation}
\sigma^2(t)=\,\left|  \psi(t)\right|  ^{2}\;.
\end{equation}
To calculate the phase $\theta(t)$ corresponding to the solution
(\ref{sigmasqrt}) of Eq. (\ref{eqausiliariak}), we look for 
a function
$F(t)$ defined by
\begin{equation}
F(t)=\frac{1}{2i\sqrt{\kappa }}\ln\frac{\psi(t)}{\psi^{\ast}%
(t)}\;,\label{F(t)}%
\end{equation}
so that
\begin{equation}
\dot{F}=\frac{1}{2i\sqrt{\kappa }}\frac{\dot{\psi}\,\psi^{\ast}-\psi
\dot{\psi}^{\ast}}{\psi\psi^{\ast}}\;.\label{dFt}%
\end{equation}
The numerator in Eq. (\ref{dFt}) can be elaborated to give%
\begin{equation}
\dot{\psi}\,\psi^{\ast}-\psi\dot{\psi}^{\ast}=2i\sqrt{AB}\sin\theta_{0}%
W_{0}=2i\sqrt{\kappa }\label{psipsistar}%
\end{equation}
where Eq. (\ref{ABsinkW0}) has been exploited. Substitution from
Eq. (\ref{psipsistar}) in Eq. (\ref{dFt}) thus yields
\begin{equation}
\dot{F}=\frac{1}{\sigma^{2}}\,\,.
\end{equation}
Then the phase $\theta(t)$ is determined by integrating (\ref{phasetheta}),
namely%
\begin{equation}
\theta(t)=\int_{t_{0}}^{t}\frac{dt^{\prime}}{\sigma^{2}(t^{\prime}%
)}=F(t)-F(t_{0})\;,\label{thetaFt}%
\end{equation}
where $F(t)$ is provided by Eq. (\ref{F(t)}). 
We shall recall this general result later.

\subsection{The Bogolubov coefficients in terms of $\sigma$}

By resorting to the operators%
\begin{equation}
Q=\sqrt{\frac{\hbar}{2\omega_{0}m_{0}}}\,\left(  \hat{a}_{0}+\hat{a}_{0}%
^{\dagger}\right)  \;,\qquad P=-i\sqrt{\frac{\hbar\omega_{0}m_{0}}{2}}\left(
\hat{a}_{0}-\hat{a}_{0}^{\dagger}\right)  \;,\label{13.1}%
\end{equation}
where $\hat{a}_{0}=\hat{a}(t_{0})$ is the (mode) annihilation operator in the
Schr\"{o}dinger representation, combining Eqs. 
(\ref{13.1}) and (\ref{4.2}) we can derive the Bogolubov transformation%
\begin{equation}
\hat{a}(t)=\mu(t)\,\,\hat{a}_{0}+\nu(t)\,\,\hat{a}_{0}^{\dagger}\label{13.2}%
\end{equation}
whose coefficients are expressed by
\begin{equation}
\mu(t)=\sqrt{\frac{m}{2\omega_{0}m_{0}}}\left(  -i\xi^{\ast}+\frac{\omega
_{0}m_{0}}{m}\sigma\right)  \;,\quad\nu(t)=\sqrt{\frac{m}{2\omega_{0}m_{0}}%
}\left(  -i\xi^{\ast}-\frac{\omega_{0}m_{0}}{m}\sigma\right)  \;,\label{13.3}%
\end{equation}
with $m_{0}=m(t_{0})$, $\omega_{0}=\omega(t_{0})$, and $\xi$ given by
Eq. (\ref{defxi}). We note that the Bogolubov transformation can be 
naturally embedded into the relations (\ref{3.1}) and (\ref{3.2}), 
where the operators $b$, $b^{\dagger}$
can be identified with $\hat{a}(t)$ and $\hat{a}^{\dagger}(t)$.
Equation (\ref{13.3}) entails
\begin{equation}
\left|  \mu\right|  ^{2}-\left|  \nu\right|  ^{2}=1\;. \label{14.1}%
\end{equation}
On the other hand, the uncertainty product can be formulated as follows
[\ref{landolfi}]:%
\begin{align}
\left(  \Delta Q\right)  \,\left(  \Delta P\right)   &  =\frac{\hbar}{2}%
\sqrt{1+4\sigma^{2}\left(  \frac{M}{2}\sigma-\dot{\sigma}\right)  ^{2}%
}=\nonumber\\
&  =\frac{\hbar}{2}\,\left|  \mu(t)-\nu(t)\right|  \,\,\left|  \mu
(t)+\nu(t)\right|  \geq\frac{\hbar}{2}\;. \label{14.2}%
\end{align}
The uncertainty formula (\ref{14.2}) is closely related to the concept of
coherent states for the generalized oscillators. Such coherent states were
constructed by Hartley and Ray in 1982 [\ref{hartley}] taking account of
the Lewis-Riesenfield theory [\ref{lewis}]. These states share all 
the features of
the coherent states of the conventional (time-independent) oscillator except
that of the uncertainty formula, in the sense that the product $\left(  \Delta
Q\right)  \,\left(  \Delta P\right)  $ turns out to be not minimum. A few
years later, Pedrosa showed that the coherent states devised by Hartley and
Ray are equivalent to squeezed states [\ref{pedrosa}].

\section{Classical view and quantum theory generation of gravitational 
waves via the Kanai-Caldirola oscillator}

\setcounter{equation}{0}

In this section we shall develop a model of propagation of 
gravitational waves based on the application of the
auxiliary equation (\ref{eqausiliaria}) for the function $\sigma(\eta)$, in
terms of which the Bogolubov coefficients can be built-up. The Bogolubov
transformation is a basic concept in the theory of particle creation in
external fields. The created particles do exist in squeezed quantum states
[\ref{grishchuk}]. According to [\ref{grishchuk}], relic gravitons created
form zero-point quantum fluctuations during cosmological evolution should now
be in strongly squeezed states. In this context the generation of
gravitational waves is of fundamental importance.

The theory of generation of gravitational waves in the inflationary 
universe scenario is based on the action [11]
\begin{equation}
S=\frac{1}{16\pi G}\int f(R)\,\sqrt{-g}\,d^{4}x\;, \label{SfR}%
\end{equation}
where $f(R)$ is an arbitrary function of the scalar curvature $R$. The theory
defined by the above action is conformally equivalent to a pure Einstein
theory with scalar-field matter. In linear theory, the gravitational waves
decouple from the matter field, so that the main problem is to fix the
background model and to desume the relation between the conformal metric.

Starting from Eq. (\ref{SfR}), by varying the action 
with respect to the gravitational perturbation 
field $h_{j}^{i}$, the equation of motion
\begin{equation}
h^{\prime\prime}+2\frac{\tilde{a}^{\prime}}{\tilde{a}}\,h^{\prime}+
(2K-\Delta)\,h=0
\end{equation}
is obtained, where prime denotes $\frac{d}{d\eta}$, 
$\tilde{a}(\eta)=\sqrt{\frac{\partial f}{\partial R}}\,
a(\eta)$,    $h=h(\eta,\overrightarrow{x})$ is  
each component of $h_{j}^{i}$, $\Delta $ stands for the 
Laplace-Beltrami operator, and $K$ means the space curvature.
By separating
in $h(\eta,\overrightarrow{x})$ the dependence on $\eta$ from the 
dependence on 
$\overrightarrow{x}$, we can write $h\sim h_{0}(\eta)\,h_{1}%
(\overrightarrow{x})$, so that
\begin{equation}
\left[  \Delta+(n^{2}-K)\right]  \,h_{1}(\overrightarrow{x})=0\;,
\end{equation}
and
\begin{equation}
h_{0}^{\prime\prime}+2\frac{\tilde{a}^{\prime}}{\tilde{a}}
\,h_{0}^{\prime}+(n^{2}+K)\,h_{0}=0\;. \label{eqh0}%
\end{equation}
Equation (\ref{eqh0}) can be applied to describe the evolution of
gravitational waves in any state of the evolution of the universe, even when
resorting to higher derivative theories of gravity [\ref{mukhanov}]. The
elimination of the first derivative in Eq. (\ref{eqh0}) leads to the 
equation (called master equation in [\ref{grishchuk}])
\begin{equation}
y^{\prime\prime}+[(n^{2}+K)-V(\eta)]\,\,y=0\,\,,
\label{yddotnKV}
\end{equation}
with 
\[
V(\eta)=\frac{a''}{a} \quad
,\quad y(\eta)=\frac{a(\eta)}{a(\eta_0)}\,h_0(\eta)\,\,.
\]
We remark that Eq. (\ref{eqh0}) can be regarded as the equation of
motion of an oscillator with time-dependent mass $m$ and constant frequency
$\bar{\omega}$,
\begin{equation}
q^{\prime\prime}+\frac{m^{\prime}}{m}q^{\prime}+
\bar{\omega}^2 q=0\,\,,
\label{qmomega}
\end{equation}
which is described by the Hamiltonian
\begin{equation}
H=\frac{p^{2}}{2m}+\frac{m\bar{\omega}^{2}}{2}q^{2}\;. \label{HKC}%
\end{equation}
The quantum theory of gravitational waves is therefore equivalent to the
quantum theory of 
the Kanai-Caldirola oscillator [\ref{kanai},\ref{caldirola}]. 
The formal analogy is realized upon the identification:
\begin{equation}
m=a^{2}\quad,\quad\bar{\omega}^{2}=n^{2}+K\quad,\quad q\sim h_{0}\;.
\end{equation}

\subsection{Exact solution of the parametrically excited oscillator and its
associated auxiliary equation}

We are mainly interested in the period under which the universe
accelerates, namely in its \textit{inflationary} stage. Inflation is defined
to be a period of accelerating expansion. During such a stage, the universe
expands adiabatically and the Friedmann equations can be exploited
[\ref{mukhanov}]. The prototype of the models of inflationary cosmology is
based on the de Sitter spacetime, which is a very interesting case concerned
with the constant Hubble rate and the scale factor given by
\begin{equation}
a(\eta)=-\frac{1}{H_{0}\eta}\quad, \label{adS}%
\end{equation}
where $\eta<0$ and $H_{0}$ denotes the Hubble constant. 
In the de Sitter case, Eq. (\ref{yddotnKV}) 
takes the form of Eq. (\ref{tdoy}), 
with $\Omega^{2}(\eta)=n^{2}-\frac{2}{\eta^{2}}$. 
It admits the general solution
\begin{equation}
y=\sqrt{-n\eta}\,\left[  \kappa_{1}J_{\frac{3}{2}}(-n\eta)+\kappa_{2}%
J_{-\frac{3}{2}}(-n\eta)\right]  \;, \label{5.1.4}%
\end{equation}
where $J_{\frac{3}{2}}$, $J_{-\frac{3}{2}}$ are Bessel functions of the first
kind and the arbitrary constants $\kappa_{1}$ and
$\kappa_{2}$ are determined once the initial conditions are imposed.
Then from Eq.(\ref{sigmasqrt}) we infer that Eq. (\ref{eqausiliariak}), 
which now reads
\begin{equation}
\sigma^{\prime\prime}+\left(  n^{2}-\frac{2}{\eta^{2}}\right)  
\,\sigma =\frac{\kappa}{\sigma^{3}}\; ,\label{5.1.5}%
\end{equation}
 is exactly solved by
\begin{equation}
\sigma=\sqrt{-n\eta}\left[  AJ_{\frac{3}{2}}^{2}(-n\eta)+BJ_{-\frac{3}{2}}%
^{2}(-n\eta)+2CJ_{\frac{3}{2}}(-n\eta)\,\,J_{-\frac{3}{2}}(-n\eta)\right]
^{\frac{1}{2}}\;,\label{5.1.6}%
\end{equation}
where constants $A,B,C$ satisfy the condition (\ref{eqABCW0}).

\subsubsection{The phase of the auxiliary field $\sigma(t)$}

As we have shown in Subsection 2.2, to calculate the phase $\theta$
corresponding to the solution (\ref{5.1.6}) of Eq. (\ref{5.1.5}) it is
convenient to introduce the function%
\begin{equation}
\psi(-n\eta)=\sqrt{-n\eta}\left[  \sqrt{A}\,e^{i\alpha}\,J_{\frac{3}{2}%
}(-n\eta)-\sqrt{B}\,e^{i\beta}\,\,J_{-\frac{3}{2}}(-n\eta)\right]  \;,
\label{5.1.7}%
\end{equation}
where $\alpha,\beta$ are real numbers. Hence the auxiliary field can be
expressed via $\sigma^{2}=\left|  \psi\right|  ^{2}$ provided that
\begin{equation}
AB\sin^{2}\theta_{0}=\frac{\kappa}{W_{0}^2}\;, \label{5.1.9}%
\end{equation}
where $\theta_{0}=\alpha-\beta$ and $W_{0}=W_{0}(\sqrt{-n\eta}J_{\frac{3}{2}%
}(-n\eta),\sqrt{-n\eta}J_{\frac{3}{2}}(-n\eta)\,)$. Therefore it is an easy
matter to see that in the case under consideration the phase $\theta$
determined by integrating Eq. (\ref{thetaFt}) is given by ($z=-n\eta$)%
\begin{equation}
\theta(\eta)=-n\int_{z_{0}}^{z}\frac{dz^{\prime}}{\sigma^{2}(z^{\prime}%
)}=-n\,[F(z)-F(z_{0})]\;, \label{5.2.6}%
\end{equation}
where the function $F(z)$ is defined by
\begin{equation}
F(z)=\frac{1}{2i\sqrt{k}}\ln\frac{\psi(z)}{\psi^{\ast}(z)}\; \label{5.2.1}%
\end{equation}
and $\psi$ is given by Eq. (\ref{5.1.7}).

\subsection{On the phase of gravitational waves}

A deep discussion on the phase of gravitational waves is contained in
[\ref{kin}], where this topic is dwelt on both at the
classical and quantum level.

Here we confine ourselves to tackle the problem classically. The study of the
phase of gravitational waves by a quantum point of view will be 
done elsewhere.

For our purpose, first we observe that the Bessel functions $J_{\frac{3}{2}}$
and $J_{-\frac{3}{2}}$ 
can be explicitly written as follows [\ref{gradshteyn}]%
\[
J_{\frac{3}{2}}(z)=\sqrt{\frac{2}{\pi z}}\,\left(  \frac{\sin z}{z}-\cos
z\right)  \quad,\quad J_{-\frac{3}{2}}(z)=-\sqrt{\frac{2}{\pi z}}\,\left(
\sin z+\frac{\cos z}{z}\right)  \;.
\]
Then, the auxiliary field (\ref{sigmasqrt}) can be written as%
\begin{align}
\sigma(z)  &  =\sqrt{\frac{2}{\pi}}\,\left[  A\left(  \frac{\sin z}{z}-\cos
z\right)  ^{2}+\,\,\,\right. \nonumber\\
&  \left.  \,\,\,\,\ \,+B\left(  \sin z+\frac{\cos z}{z}\right)
^{2}-2C\left(  \frac{\sin z}{z}-\cos z\right)  \left(  \sin z+\frac{\cos z}%
{z}\right)  \right]  ^{\frac{1}{2}}\,, \label{sigmaGW}%
\end{align}
where $y_{1}=\sqrt{z}J_{\frac{3}{2}}$, $y_{2}=\sqrt{z}J_{-\frac{3}{2}}$ and
the condition (\ref{eqABCW0}) is understood.

Now we shall see that the phase 
$\theta_{GW}$ of the primordial gravitational
waves can be obtained by Eq. (\ref{sigmaGW}) under the choice
\begin{equation}
A=B\neq0\;,\quad C=0\;, \label{ABCGW}%
\end{equation}
and assuming asymptotically negative values of the conformal time. In doing
so, Eqs. (\ref{5.2.6}), (\ref{sigmaGW}) provide
\begin{equation}
\sigma_{in}^{2}\sim\frac{2A}{\pi} \qquad,
\qquad
\theta_{GW}\equiv\theta_{in}=\frac{\pi}{2A}\left(  z-z_{0}\right)  
\; \; ,
\label{thetaGW}%
\end{equation}
with 
\begin{equation}
A=\frac{\pi\sqrt{\kappa}}{2n} \; .
 \label{AGW}
\end{equation}
In the case $n^{2}\gg \left|  V(\eta)\right| $, 
the high frequency waves,
\textit{e.g.} the solutions of the equation
\begin{equation}
y^{\prime\prime}+n^{2}y=0\;,
\end{equation}
correspond to the following behavior of the gravitational perturbation field
$h$: 
\begin{equation}
h(\eta)=\frac{1}{a}\sin\left(  n\eta+\rho\right)
\end{equation}
($\rho$ is an arbitrary phase). In an expanding universe, the amplitude $h$ of
the waves decreases adiabatically for all $\eta$. The result represented by
the calculation of the phase (\ref{thetaGW}) of relic gravitational waves
suggests one to interpret formula (\ref{5.2.6}) as the phase of gravitational
waves not only in the case of the primordial cosmological scenario formally
corresponding to $\eta\rightarrow-\infty$. Anyway, this subject, which goes
beyond the scope of the present paper, deserves further investigation.
Here we recall only that the general solution of Eq. (\ref{tdoy}), which
holds in the case of the de Sitter cosmological model, can also 
be written as
\begin{equation}
y=\frac{c}{\sqrt{\kappa}}\sigma(\eta)\,\cos\left[  \sqrt{\kappa}\theta
(\eta)+\delta\right]
\end{equation}
where $\theta$ is given by Eq. 
(\ref{5.2.6}), $c$ is a Noether invariant of Eq.
(\ref{tdoy}), and $\delta$ is an arbitrary constant. 
Hence for any conformal
time $\eta$ in the interval $\left(  -\infty,0\right)  $, the amplitude $h$ of
the gravitational perturbation field can be expressed by
\[
h\sim\frac{1}{a}\sigma(\eta)\,\sin\left[  \sqrt{\kappa}\theta(\eta
)+const\right]  \;.
\]

\subsection{Quantum gravitational waves: Theory in terms of the auxiliary
field $\sigma(\eta)$}

The approach to the study of gravitational waves we present in
this paper is developed starting from the classical 
Kanai-Caldirola Hamiltonian
[\ref{kanai},\ref{caldirola}], Eq. (\ref{HKC}). 
As we have already pointed out, 
Eq. (\ref{eqh0}) can be regarded, in fact, as
the equation of motion of an oscillator with time-dependent mass $m=a^{2}$ 
and constant frequency $\bar{\omega}=\sqrt{n^{2}+K}$ (see Eq. 
(\ref{qmomega})) described by the Hamiltonian
(\ref{HKC}).
So, we make the fundamental identification of the whole temporal part 
$h_{0}$ of the metric fluctuation amplitude as the basic ''coordinate'' 
variable to quantize as such. (Recall that our procedure for 
the quantization of gravitational waves is based on the identification
of $y=a\,h_{0}$ as the variable to quantize). As a consequence, the quantum
theory of gravitational waves turns out to be completely equivalent to the
quantum theory of the Kanai-Caldirola oscillator which can be described by 
the quantum version of the Hamiltonian (\ref{HKC}). On the
grounds of what we learned in Section 2, the above identification
suggests a route which can be successfully pursued whenever we are
interested in the characterization of physical effects (quantum decoherence,
squeezing, particle production etc.) emerging from the study of inflationary
models in the early universe. Section 4 will be devoted to a preliminary
exploration of the effectiveness of the idea in the context of expanding
universe cosmological models.

\section{ Applications}
\setcounter{equation}{0}

By taking full advantage of the formalism introduced in Section 2, we are in
the position to study the dynamical system of the cosmological interest which
are described by time-dependent oscillators. In doing so, a key point is the
characterization of constants $A,B,C$ in Eq. (\ref{sigmasqrt}). It is
concerned with the initial (and boundary) conditions. 
All dynamical aspects under
time evolution are enclosed into the function $\sigma$, which obeys the second
order nonlinear differential equation (\ref{eqausiliaria}). A general
condition on $A,B,C$ is provided by Eq. (\ref{eqABCW0}). It is not enough,
however. Specification of the value and the first time derivative of $\sigma$
at fixed time is thereby needed. Another condition is associated with the
requirement that at the initial time $\eta_{i}$ the time-dependent
annihilation and creation operators, derived from Eq. (\ref{4.2}), go into the
standard Dirac-like form, Eq. (\ref{13.1}). As for the final condition, it
is helpful to reveal that in most cases we want the state at initial time to
correspond to a vacuum state. This can be achieved easily under the
minimization requirement for $E_{NM}=\frac{\hbar}{2}\left(  \frac{M}{2}%
\sigma-\dot{\sigma}\right)  ^{2}$. Indeed, since it provides a measure of the
decoherence at the time $\eta$, it has to be vanishing when referring to a
vacuum state at the initial time $\eta=\eta_{i}$. Under these circumstances,
the whole set of initial conditions for $\sigma$ is given by%
\begin{equation}
\left\{
\begin{array}
[c]{l}%
AB-C^{2}=\frac{\kappa}{W_{0}^{2}}\;,\\
\sigma(t_{0})-\left(  4\omega^{2}\right)  ^{-1/4}=0\;,\\
\dot{\sigma}(t_{0})-\frac{M(t_{0})}{2}\sigma(t_{0})=0\;.
\end{array}
\right.
\end{equation}
In a cosmological framework of the FRW type, above system is translated into%
\begin{equation}
\left\{
\begin{array}
[c]{l}%
AB-C^{2}=\frac{\kappa}{W_{0}^{2}}\;,\\
\sigma(\eta_{i})=[4(n^{2}+K)]^{-1/4}\;,\\
\sigma^{\prime}\left(  \eta_{i}\right)  -\frac{a^{\prime}(\eta_{i})}%
{\sqrt{2 n}a(\eta_{i})}=0\;.
\end{array}
\right.  \label{sysABCin}%
\end{equation}

In order to proceed with concrete analysis, it is very customary to resort to
the spatially flat inflationary model based on the de Sitter metric.
However, a more general and realistic description of the
inflation may be provided by a quasi-de Sitter spacetime (see \textit{e.g.}
[\ref{riotto}]). In this case, the Hubble rate is not exactly constant but,
rather, it weakly conformal changes with time according to $\tilde{H}^{\prime
}=-\epsilon a^{2}\tilde{H}^{2}$ (that is, $aa^{\prime\prime}=(2-\epsilon
)\,a^{4}\tilde{H}^{2}\,=(2-\epsilon)\,a^{\prime2}$) where $\epsilon$ is a
constant parameter. When $\epsilon$ vanishes one gets just the ordinary de
Sitter spacetime. For small values of $\epsilon$, a quasi-de Sitter spacetime
is associated with the scale factor%
\begin{equation}
a(\eta)=\frac{-1}{\tilde{H}(1-\epsilon)\eta}\;
\end{equation}
($\eta<0$). In the quasi-de Sitter spatially flat scenario, Eq. (\ref{tdoy}%
) reads%
\begin{equation}
y^{\prime\prime}+\left[  n^{2}-\frac{(2+3\epsilon)}{(1-\epsilon)^{2}\eta^{2}%
}\right]  \,y=0
\end{equation}
and can be solved in terms of Bessel functions. Precisely, one has the two
independent solutions
\begin{equation}
y_{1}=\sqrt{-n\eta}J_{\nu}\left(  -n\eta\right)  \quad,\quad y_{2}%
=\sqrt{-n\eta}\,Y_{\nu}\left(  -n\eta\right)  \label{y12ddS}%
\end{equation}
where $\nu=\sqrt{\frac{1}{4}+\frac{(2+3\epsilon)}{(\epsilon-1)^{2}}}$ . The
procedure outlined in the previous 
sections can be applied and we are led to the
introduction of the basic function%
\begin{align}
\sigma &  =\left(  Ay_{1}^{2}+By_{2}^{2}+2Cy_{1}y_{2}\right)  ^{\frac{1}{2}%
}=\nonumber\\
&  =\sqrt{-n\eta}\left\{  AJ_{\nu}^{2}\left(  -n\eta\right)  +BY_{\nu}%
^{2}\left(  -n\eta\right)  +2CJ_{\nu}\left(  -n\eta\right)  Y_{\nu}\left(
-n\eta\right)  \right\}  ^{\frac{1}{2}}%
\end{align}
where $A,B,C$ are determined by means of the system (\ref{sysABCin}),
$\eta_{i}$ denoting the conformal time of 
the beginning of the inflation. Once we
are interested in a situation in which the system started very far in the past
in a vacuum state, the Bessel function expansions
\begin{align*}
J_{\nu}(-n\eta) &  \sim\sqrt{-\frac{2}{\pi n\eta}}\left[  \cos(-n\eta
-\frac{\nu}{2}\pi-\frac{\pi}{4})+O\left(  \frac{1}{n\eta}\right)  \right]
\,\,,\,\,\\
Y_{\nu}(-n\eta) &  \sim\sqrt{-\frac{2}{\pi n\eta}}\left[  \sin(-n\eta
-\frac{\nu}{2}\pi-\frac{\pi}{4})+O\left(  \frac{1}{n\eta}\right)  \right]
\,\,,
\end{align*}
for $\nu$ fixed and $n\eta\rightarrow-\infty$ assists us in finding 
suitable constants 
$A,B,C$. By taking arbitrary asymptotically negative initial times, the
leading terms of Bessel functions $J_{\nu}$, $Y_{\nu}$ give rise to the
following behavior for the function:
\[
\sigma(\eta)=\sqrt{\frac{2}{\pi}}\left\{  A+(B-A)\sin^{2}(-n\eta-\frac{\nu}%
{2}\pi-\frac{\pi}{4})+C\sin(-2n\eta-\nu\pi-\frac{\pi}{2})+O\left(  \frac
{1}{n\eta}\right)  \right\}  ^{\frac{1}{2}}\,\,.
\]
Once the limit $\eta_{i}<<0$ is concerned a natural choice is given by
$A=B=\frac{\pi}{4n}$, $C=0$ (recall that we already found this result 
for the case $\nu=3/2$ associated with the standard de Sitter
metric background). So we obtain%
\begin{equation}
\sigma(\eta)=\sqrt{-\frac{\pi}{4}\eta}\,\left\{  J_{\nu}^{2}\left(
-n\eta\right)  +Y_{\nu}^{2}\left(  -n\eta\right)  \right\}  ^{\frac{1}{2}%
}=\sqrt{-\frac{\pi}{4}\eta}\,\left|  H_{\nu}^{1}(-n\eta)\right|
\,\,.\label{sigmaqdS}%
\end{equation}
In the light of our previous results, the decoherence energy $E_{NM}$ at the
time $\eta$ of gravitational waves in a quasi-de Sitter model of inflation can
be evaluated by inserting (\ref{sigmaqdS}) into formula (\ref{10.2}). It then
results%
\[
E_{NM}=\frac{\hbar}{2}\left[  \sigma^{\prime}-\frac{a^{\prime}(\eta)}{a(\eta
)}\sigma\right]  ^{2}=\frac{\hbar}{2}\left[  \sigma^{\prime}+\frac{\sigma
}{(1-\epsilon)\,\eta}\right]  ^{2}\,\,
\]
where%
\begin{equation}
\sigma^{\prime}+\frac{\sigma}{(1-\epsilon)\,\eta}=\sqrt{\frac{\pi}{4}}\left\{
\frac{n}{2}\frac{\sqrt{-\eta}}{\left|  H_{\nu}^{1}(-n\eta)\right|  }\left(
H_{\nu}^{1\ast}H_{\nu+1}+c.c\right)  -\left(  \nu+\frac{3-\epsilon
}{2(1-\epsilon)}\right)  \frac{\left|  H_{\nu}^{1}(-n\eta)\right|  }%
{\sqrt{-\eta}}\right\}  \,\,.\label{sdotseta}%
\end{equation}
Moreover, since%
\begin{align}
\mu(\eta) &  =\sqrt{\frac{a^{2}(\eta)}{2n\,a^{2}(\eta_{i})}}\left\{  \left[
\frac{1}{2\sigma}+\frac{n\,a^{2}(\eta_{i})}{a^{2}(\eta)}\sigma(\eta)\right]
-i\left[  \sigma^{\prime}-\frac{a^{\prime}}{a}\sigma\right]  \right\}
\;,\quad\\
\nu(\eta) &  =\sqrt{\frac{a^{2}(\eta)}{2n\,a^{2}(\eta_{i})}}\left\{  \left[
\frac{1}{2\sigma}-\frac{n\,a^{2}(\eta_{i})}{a^{2}(\eta)}\sigma(\eta)\right]
-i\left[  \sigma^{\prime}-\frac{a^{\prime}}{a}\sigma\right]  \right\}  \;
\end{align}
at an arbitrary time $\eta$ the Bogolubov coefficients are given by%
\begin{align}
\mu(\eta) &  =\sqrt{\frac{1}{2n}\,\left(  \frac{\,\eta{}_{i}}{\eta{}}\right)
^{\frac{2}{1-\epsilon}}}\left\{  \left[  \frac{1}{2\sigma}+\left(
\frac{\,\eta{}}{\eta_{i}{}}\right)  ^{\frac{2}{1-\epsilon}}n\sigma\right]
-i\left[  \sigma^{\prime}+\frac{\sigma}{(1-\epsilon)\,\eta}\right]  \right\}
\;,\quad\\
\nu(\eta) &  =\sqrt{\frac{1}{2n}\,\left(  \frac{\,\eta{}_{i}}{\eta{}}\right)
^{\frac{2}{1-\epsilon}}}\left\{  \left[  \frac{1}{2\sigma}-\left(
\frac{\,\eta{}}{\eta_{i}{}}\right)  ^{\frac{2}{1-\epsilon}}n\sigma\right]
\ -i\left[  \sigma^{\prime}+\frac{\sigma}{(1-\epsilon)\,\eta}\right]
\right\}  \;
\end{align}
with $\sigma$ and $\sigma^{\prime}+\frac{\sigma}{(1-\epsilon)\,\eta}$
furnished by Eqs. (\ref{sigmaqdS}) and (\ref{sdotseta}), respectively. 
Finally, the
phase $\theta$ can be evaluated. Due to Eq. (\ref{thetaFt}), we get%
\begin{equation}
\theta\left(  \eta\right)  =\left.  -i \ln\frac
{\psi(\eta)}{\psi^{\ast}(\eta)}\right|  _{\eta_{i}}^{\eta}\;,
\end{equation}
where%
\[
\psi(t)=\sqrt{\frac{\pi}{4n}}\,e^{i\alpha}\left[  \,y_{1}+iy_{2}\right]
=\sqrt{\frac{\pi}{4n}}\,e^{i\alpha}\sqrt{-n\eta}\,H_{\nu}^{1}\,\,.
\]
That is,
\[
\theta\left(  \eta\right)  =\left.  2 \,\theta_{\nu}%
^{1}\right|  _{\eta_{i}}^{\eta}%
\]
where $\theta_{\nu}^{1}$ denotes the phase of the Hankel function $H_{\nu}%
^{1}$.

It is now instructive to focus on a standard de Sitter inflation. 
In this case
$\epsilon=0$, $\nu=\frac{3}{2}$ and 
\begin{equation}
\sigma=\frac{1}{\sqrt{2n}}\sqrt{1+\frac{1}{n^{2}\eta^{2}}}\qquad.
\label{sigmadssimple}%
\end{equation}
So, the integration of $\frac{1}{\sigma^{2}}$ is straight,
\textit{i.e.}
\begin{equation}
\int^{\eta}_{\eta_i}
\frac{d\eta'}{\sigma^{2}(\eta')}=
4n^{2}\int\frac{n^{2}\eta^{2}}{1+n^{2}\eta^{2}}%
d\eta=4n^{2}\left[  \eta-\frac{\tanh^{-1}(n\eta)}{n}\right]  \,\,,
\end{equation}
up to a constant of integration. The
above two formulas provide the standard exact (and normalized) solution for
the quantum fluctuations of a generic massless scalar field during a de Sitter
inflation. Yet, it is
interesting to consider a very long inflation by letting the system
evolve towards $\eta\rightarrow0$. In this case $E_{NM}$ simply reads%
\begin{equation}
E_{NM}(\eta)=\frac{\hbar}{2}\left(  \dot{\sigma}-\frac{M}{2}\sigma\right)
^{2}=\frac{\hbar}{4}\frac{n}{1+n^{2}\eta^{2}}\,\,.
\end{equation}
Interestingly, the decoherence energy at $\eta=0$ is finite. Moreover, in the
standard de Sitter phase, from Eq. 
(\ref{sigmadssimple}) we obtain the Bogolubov coefficients
\begin{align}
\mu(\eta)  &  =\frac{1}{2}\frac{\sqrt{n^{2}\eta_{i}^{2}}}{\sqrt{1+n^{2}%
\eta^{2}}}\left\{  \left[  1+\frac{\eta^2}{\eta_{i}^{2}}
\right]  -\frac{i}{n\eta\,}\right\}  \;,\quad\\
\nu(\eta)  &  =\frac{1}{2}\frac{\sqrt{n^{2}\eta_{i}^{2}}}{\sqrt{1+n^{2}%
\eta^{2}}}\left\{  \left[  1-\frac{\eta^2}{\eta_{i}^{2}}
\right]  -\frac{i}{n\eta\,}\right\}  \;,
\end{align}
($n^2 \eta_i^2 \gg1$) which in turns imply that
\begin{equation}
\left|  \nu(\eta)\right|  ^{2}=\frac{1}{4}\left(  \frac{\eta_{i}}{\eta}%
-\frac{\eta}{\eta_{i}}\right)  ^{2}  \label{nu2ds}%
\end{equation}
particles are created out the vacuum at the time $\eta$.

In more refined studies of cosmological effects in the expanding Universe, it
turns out to be useful to introduce a cosmological model which allows one to
take into account different evolutionary phases of the universe. Once the
model has been specified and the equation (\ref{tdoy}) solved, one can get
an insight into physical effects associated with different cosmological
stages. For instance, one can consider a simple cosmological model which
includes the inflationary \textit{(i)}, radiation-dominated \textit{(e)} and
matter-dominated \textit{(m)} epochs [\ref{kolb}]. The scale factor has the
following dependence on the conformal time:
\begin{equation}
\left\{
\begin{array}
[c]{ll}%
a_{i}(\eta)=-\frac{1}{H_{0}\eta}\,\,, & \qquad\quad\eta_{i}\leq\eta\leq
\eta_{e}<0 \,\, , \\
a_{e}(\eta)=\frac{\left(  \eta-2\eta_{e}\right)  }{H_{0}\,\eta_{e}^{2}}\,\,, &
\qquad\quad\eta_{e}\leq\eta\leq\eta_{m}\,\,,\\
a_{m}(\eta)=\frac{\left(  \eta+\eta_{m}-4\eta_{e}\right)  ^{2}}{4H_{0}\eta
_{i}^{2}\,(\eta_{2}-2\eta_{1})}\,\,, & \qquad\quad\eta\geq\eta_{m}\,\, ,
\end{array}
\right.  \label{aeta3fasi}%
\end{equation}
where $H_{0}$ denotes the Hubble constant at the inflationary stage, and
$\eta_{i}$ represents the beginning of the expansion. In order to determine
the $\eta$-dependent amplitude $h_{n}$ for each epoch we have to solve Eq.
(\ref{tdoy}) with the corresponding varying frequencies, namely
\begin{equation}
\Omega_{i}=\left(  n^{2}-\frac{2}{\eta^{2}}\right)  ^{\frac{1}{2}}%
\quad,\quad\Omega_{e}=n\quad,\quad\Omega_{m}=\left[  n^{2}-\frac{2}{\left(
\eta+\eta_{m}-4\eta_{e}\right)  ^{2}}\right]  ^{\frac{1}{2}}\;.
\label{5.7.7.2}
\end{equation}
Notice that the frequency $\Omega_{e}$ is constant, while $\Omega_{i}$ and
$\Omega_{m}$ are varying with the same temporal dependence. The frequency
$\Omega_{i}$ characterizes the de Sitter era. The related equation of motion
(\ref{tdoy}) has already been solved. Taking care about
matching data at $\eta_{e}$ and $\eta_{m}$ is needed for the knowledge of the
complete form of $\sigma$. The associated $\sigma$'s and their derivatives
have to join continuosly at $\eta_{e}$ and $\eta_{m}$, in fact. This step is
needed to obtain all the physical information implied in formulas for the
Bogolubov coefficients, the decoherence energy, the gravitational phase, and
squeezing. Having in mind our previous discussion, employing the model by
considering a quasi-de Sitter phase is straight.

In general, the vacuum expectation value of the number operator and the 
other quantities of the physical interest vary slowly with time if the 
expansion rate becomes arbitrarily slow. In case the espansion is 
stopped one should be able to recover time-independent Dirac operators. 
However, the circumstance
does not mean that Bogolubov coefficients trivialize. This is because loss of
coherence previously occurred due to the expansion dynamics. A typical
situation may be that $\dot{\sigma}$ goes to zero but $\sigma$ does not.
$E_{NM}$ goes to zero, indicating that when expansion is stopped the
time-dependent gravitational pumping stops as well and there is no further
decoherence. If expansion stops from time $\eta_{a}$ to time $\eta_{b}$, then
$\forall\eta\in\lbrack\eta_{a},\eta_{b}]$ one gets
\begin{align}
\mu(\eta) &  =\sqrt{\frac{a^{2}(\eta_{a})}{2na^{2}(\eta_{i})}}\left[  \frac
{1}{2\sigma(\eta_{a})}+n\frac{a^{2}(\eta_{i})}{a^{2}(\eta_{a})}\,\sigma
(\eta_{a})\right]  \;,\quad\\
\nu(\eta) &  =\sqrt{\frac{a^{2}(\eta_{a})}{2na^{2}(\eta_{i})}}\left(  \frac
{1}{2\sigma(\eta_{a})}-n\frac{a^{2}(\eta_{i})}{a^{2}(\eta_{a})}\,\sigma
(\eta_{a})\right)  \;,
\end{align}
which implies%
\begin{equation}
\left|  \mu(\eta)\right|  ^{2}=\frac{1}{2}\left[  \frac{a^{2}(\eta_{a}%
)}{4na^{2}(\eta_{i})\,\sigma^{2}(\eta_{a})}+n\frac{a^{2}(\eta_{i})}{a^{2}%
(\eta_{a})}\,\sigma^{2}(\eta_{a}) +1 \right]  \,\,.
\end{equation}

To get a more clear insight into the results achieved in this section, 
a few comments are in order. Specifically, our formula for the 
decoherence energy associated with the dynamical evolution of 
the gravitational fluctuation modes on a background of the 
FRW type, is expressed in a very compact form 
in terms of the auxiliary field $\sigma (\eta)$ and the scale factor 
$a(\eta)$. On the other hand, the decoherence energy plays an 
essential role in the relationships for the Bogolubov coefficients. 
This aspect makes explicit how the energy lost owing to the decoherence 
effect may be exploited to excite the vacuum state of the model under 
consideration. Moreover, in our framework this mechanism would be 
quantified in a compact way by means of the formula
\begin{equation}
\label{NU}
|{\nu}(\eta)|^{2}=\frac{a^2(\eta)}{2 n a^2(\eta_i)}
\left[ \left( \frac{1}{2\sigma}-\frac{n a^2(\eta_i)}{a^2(\eta)} 
\sigma \right)^{2} 
+ \frac{2}{\hbar}E_{NM} \right],
\end{equation}
where $E_{NM}$ is the decoherence energy $(m=a^{2})$.
With respect to other works, in our paper the role of the decoherence 
energy is made manifest. Furthermore, we observe that, remarkably, 
the auxiliary field $\sigma $ is nothing but the time-dependent 
amplitude of the mode solutions to  Eq. (\ref{tdoy}) for the 
redshifted gravitational field fluctuations.

\section{Concluding remarks}
\setcounter{equation}{0}

The main results achieved in this paper have been presented and widely
discussed in the Introduction. Therefore we shall conclude 
by making some final comments concerning
challenging perspectives which should be dwelt upon in future developments.
The evolution equation of a mode with comoving wavenumber $n$
reduces to the harmonic oscillator equation with time-dependent mass and
constant frequency. The approximation behind computations leading to the
result actually are applicable only to the infrared region. On general
grounds, one therefore expects that predictions for observables may depend
sensitively on the physics on the length scales smaller than the Planck one.
In order to take into account trans-Planckian physics, it has been recently
suggested to make use of effective dispersion relations (see \textit{e.g.}
[\ref{martin}]). The linear dispersion relation is thus replaced by a
nonlinear one, $n_{eff}^{2}=a^{2}(\eta)\,F^{2}(n/a)$ , where $F(n/a)$ is an
arbitrary function required to behave linearly whenever $n/a$ ($=k$) is below
a certain threshold. A time dependent dispersion relation thus enters in the
matter. As a consequence, the underlying dynamical model turns out to be that
of the harmonic oscillator with both the mass $m$ (=$a^{2}$) and the frequency
$\omega$ ($=n_{eff}$) depending on time. In principle, its quantization can
still be pursued by resorting to the formalism of Section 2 and it will be
studied in detail elsewhere. Nevertheless, a comment is in order. In the
light of the 
discussion in [\ref{landolfi}], one might wonder, in fact, on whether
or not in the cosmological framework the minimum uncertainty criterium can be
satisfied under time evolution for some physically reasonable function $F$. It
is straightforwardly seen that this is not the case, generally speaking.
Indeed, in the cosmological framework the criterium reads as $a^{2}%
n_{eff}=const$ and implies a purely cubic function $F$, say $F=\alpha_{0}%
n^{3}/a^{3}$. As a consequence, the uncertainty relation can be minimized only
approximately. It is worth noting that this happens in the large wavenumbers
limit of a special case of the generalized Corley-Jacobson dispersion relation
introduced in [\ref{martin}] (see Eq. (22) in [\ref{martin}]).%

\section*{References}%

\begin{enumerate}
\item \label{grishchuk}L.P. Grishchuk and Y.V. Sidorov, Phys. Rev.
\textbf{D42}, 3413 (1990) .

\item \label{stoler}D. Stoler, Phys. Rev. \textbf{D1}, 3217 (1970);
\textbf{D4}, 1925 (1971).

\item \label{yuen}H.P. Yuen, Phys. Rev. \textbf{A13}, 2226 (1976) .

\item \label{schumaker} B.L. Schumaker, Phys. Rep. \textbf{135}, 317 (1986).

\item \label{birrel}N.D. Birrel and P.C.W. Davies, \textit {Quantum fields in
curved space}, Cambridge Univ. Press., Cambridge (1994) .

\item \label{grishchuk2}L.P. Grishchuk, Class. Quantum Grav.
\textbf{10}, 2449 (1993).

\item L.P. Grishchuk and Y.V. Sidorov, Class. Quant. Grav. \textbf{6},
L161 (1989).

\item L.P. Grishchuk and Y.V. Sidorov, Phys. 
Rev. \textbf{D42}, 3413 (1990).

\item L.P. Grishchuk, Class. Quant. Grav. \textbf{10}, 2449 (1990).

\item \label{bose}S. Bose and L.P. Grishchuk, Phys. Rev. 
\textbf{D66}, 043529 (2002).

\item \label{mukhanov}V.F. Mukhanov, H.A. Feldman and R.H. Brandenberger,
Phys. Rep. \textbf{215}, 203 (1992).

\item \label{maggiore}M. Maggiore, Phys. Rep. \textbf{331}, 283 (2000).

\item \label{profilo}G. Profilo and G. Soliani, Phys. Rev. \textbf{A44}, 2057
(1991) .

\item \label{landolfi}G. Landolfi, G. Ruggeri and G. Soliani, 
\textit{Amplitude and phase of time dependent Hamiltonian systems 
under the minimum uncertainty condition}, preprint, quant-ph/0307109 (2003).

\item \label{hollerhorst}J.H. Hollenhorst, Phys. Rev.\textbf{ D19}, 1669 (1979).

\item \label{lewis}H.R. Lewis,Jr. and W.B. Riesenfeld, J. Math. Phys.
\textbf{10}, 1458 (1969).

\item \label{eliezer}C.J. Eliezer and A. Gray, SIAM J. Appl. Math.
\textbf{30}, 463 (1976).

\item \label{ermakov}V.P. Ermakov, Universiteskiya Izvestiya Kiev 
\textbf{20}, 1 (1880).

\item \label{pinney}E. Pinney, Proc. Am. Math. Soc. \textbf{1}, 681 (1950).

\item \label{goff}S. Goff and D.F. St. Mary, J. Math. Anal. Appl.
\textbf{140}, 95 (1989).

\item \label{hartley}J.G. Hartley and J.R. Ray, Phys. Rev. \textbf{D25}, 382
(1982) .

\item \label{pedrosa}I.A. Pedrosa, Phys. Rev. \textbf{D36}, 1279 (1987).

\item \label{kanai}E. Kanai, Prog. Theor. Phys. \textbf{3}, 440 (1948).

\item \label{caldirola}L. Caldirola, Nuovo Cimento \textbf{B87}, 241 (1983).

\item \label{kin}Kin-Wang Ng and A.D. Speliotopoulos, Phys. Rev. \textbf{D51},
5636 (1995).

\item \label{gradshteyn}I.S. Gradshteyn and I.M. Ryzhik, \textit{Table of
Integrals, Series and Products}, Academic Press, New York (1965).

\item \label{martin}J. Martin and R.H. Brandenberger, Phys.Rev. \textbf{D63},
123501 (2001).

\item \label{riotto}A. Riotto, \textit{Inflation and the theory of
cosmological perturbations}, Lectures delivered at the ''ICTP Summer School on
Astroparticle Physics and Cosmology'', Trieste, 17 June - 5 July 2002,
hep-ph/0210162 (2002).

\item \label{kolb}E.W. Kolb, H.S. Turner, \textit{The early universe},
Addison-Wisley, Reading, 1990.
\end{enumerate}
\end{document}